# Model-based Testing of Object-Oriented Systems

Bernhard Rumpe

IRISA-Université de Rennes 1, Campus de Beaulieu, Rennes, France
and Software & Systems Engineering, TU München, Germany

This paper discusses a model-based approach to testing as a vital part of software development. It argues that an approach using models as central development artifact needs to be added to the portfolio of software engineering techniques, to further increase efficiency and flexibility of the development as well as quality and reusability of results. Then test case modeling is examined in depth and related to an evolutionary approach to model transformation. A number of test patterns is proposed that have proven helpful to the design of testable object-oriented systems. In contrast to other approaches, this approach uses explicit models for test cases instead of trying to derive (many) test cases from a single model.

## 1 Portfolio of Software Engineering Techniques

Software has become a vital, but often invisible part of our lives. Embedded forms of software are part of almost any technical device. The average household uses several computers, and the internet and telecommunication world has considerably changed our lives. Software is used for a variety of jobs. It can be as small as a simple script or as complex as an entire operating or enterprise resource planning system.

For the near future, we can be rather sure that we will not have a single notation or process that can cover the diversity of today's development projects. Projects are too different in their application domain, size, need for reliability, time-to-market pressure, and the skills and demands of the project participants. Even the UML [OMG02], which is regarded as a de-facto standard, is seen as a family of languages rather than a single notation and by far doesn't cover all needs. This leads to an ongoing proliferation of methods, notations, principles, techniques and tools in the software engineering domain that is at least partly influenced from practical applications of Formal Methods.

On the one hand, methods like Extreme Programming [Bec99] and Agile Software Development [Coc02] even discourage the long well known distinction between analysis, design and implementation activities and abandon all documentation activities in favor of rigorous test suites. On the other hand, upcoming development tools allow to generate increasing amounts of code from UML models, thus supporting the OMG's initiative on "Model Driven Architecture" (MDA) [OMG01]. MDA's primary purpose is to decouple platform-independent models from platform-specific, technical information. This should increase the reusability of both. Code generation, however, is today focusing pretty much on the generation of the productive system. Generation of test code from models is still a side issue. In particular the question, what a good test model should look like, is to be examined in this paper.

In general, we can observe that in the foreseeable future we will have a *portfolio of software engineering techniques* that enables developers and managers to select appropriate processes and tools for their projects. Today, however, it is not quite clear which elements the portfolio should have, how they relate, when they are applicable, and what their benefits and drawbacks are. The software and systems engineering community therefore must reconsider and extend its portfolio of software engineering techniques incorporating new ideas and concepts, but also try to scientifically assess the benefits and limits of new approaches. For example:

- Lightweight projects that don't produce requirement and design documentation need intensive communications and can hardly be split into independent subprojects. Thus they don't scale up to large projects. But where are the limits? A guess is, around 10 people, but there have been larger projects reportedly "successful" [RS02].
- Formal methods have built a large body of knowledge (see for example [ABRS03,JKW03,LCCRC03]), but how can this knowledge successfully and in a goal-oriented way be applied in today's projects? A guess seems to be, formal methods apply best, if embodied in practical tools, using practical and well known notations without exposing the user directly to the formal method.
- Product reliability often need not be 100% for all developments and in the first iteration already. But how to predict reliability from project metrics and how to adapt the project to increase reliability and accuracy to the desired level while minimizing the project/product costs?

Thus in contrast to applying formal methods for verification purposes, the use of formal techniques for test case specification and metrics of test coverage does not give 100% reliability, but in practice has a much better cost/benefit ratio. Based on this observation we will in the following examine the modeling of test cases using several of the UML-like notations, arguing that this technique should be a new item in the broad portfolio of SE techniques. For this purpose, we develop our interpretation of the used UML notation in the context of test case modeling, which should give us the justification to regard the used notation as being backed up by a formal technique (without explicitly referring to that formal method).

Section 2 discusses synergies and problems of using models for a variety of activities, including programming. Section 3 establishes the general needs of successful testing strategies. In Section 4 the scenario of a model-based test approach is discussed. Sections 5 and 6 present several test patterns that are useful to make an object-oriented design testable. While Section 5 concentrates on basic test patterns, Section 6 presents more specific test patterns for distributed systems. Section 7 finally discusses the benefits of an evolutionary approach to modeling in combination with an intensive, model-based test approach. In particular, the usability of tests as invariant observations for model-transformations is explored. For sake of conceptual discussions, technical details are omitted, but can be found in [Rum03].

## 2   Modeling meets Programming

UML [OMG02] undoubtedly has become the most popular modeling language for software intensive systems used today. Models can be used for quite a variety of purposes. Besides informal sketches that are used for *communication*, e.g. by being drawn on paper and posted on a wall the most common are:
- Semi-precisely defined diagrams are used for *documentation* of that part of the requirements that is not written in plain English.
- Architecture and designs are captured and documented with models. In practice, these models are increasingly often used for *code generation*.

More sophisticated and therefore less widespread uses of models are analysis of certain quality attributes (such as message throughput, responsiveness or failure likelihood) or development of tests from models. Many UML-based tools today offer functionality to directly simulate models or generate at least parts of the code. As tool vendors work hard on continuous improvement of this feature, this means a sublanguage of UML will become a high-level programming language and modeling at this level becomes identical to programming. This raises a number of interesting questions:
- Is it critical for a modeling language to be also used as programming language? For example analysis and design models may become overloaded with details that are not of interest yet, because modelers are addicted to executability.
- Is a future version of the UML expressive enough to describe systems completely or will it be accompanied by conventional languages? How well are these integrated?
- How will the toolset of the future look like and how will it overcome round trip engineering (i.e. mapping code and diagrams in both directions)?
- What implications does an executable UML have on the development process?

In [Rum03,Rum02] we have discussed these issues and have demonstrated, how the UML in combination with Java may be used as a high-level programming language. But, UML cannot only be used for modeling the application, but more importantly for modeling tests on various levels (class, integration, and system tests) as well. For this purpose we need executable test models, as testing is in general the process of executing a program with the intention to identify faults [Mye79,Bin99]. Executable models are usually less abstract than design models, but they are still more compact and abstract than the implementation. The same holds for test models versus manually implemented tests.

One advantage of using models for test case description is that application specific parts are modeled with UML-diagrams and technical issues, such as connection to frameworks, error handling, persistence, or communication are handled by the parameterized code generator. This basically allows us to develop models that are independent of any technology and platform, as for example proposed in [SD00]. Only during the generation process platform dependent elements are added. When the technology changes, we only need to update the generator, but the application defining models as well as test models can directly be reused. This concept also directly supports the above mentioned MDA-Approach [Coc02] of the OMG. Another impor-tant advantage is that both, the production code and automatically executable tests at any level, are modeled by the same UML diagrams. Therefore, developers use a single homogeneous language to describe implementation and tests. This will enhance the availability of tests already at the beginning of the coding activities and leads to a development method similar to the "test first approach" [Bec01,LF02].

Some of the UML-models (mainly class diagrams and statecharts) are used constructively, others are used for test case definition (mainly OCL, sequence and enhanced object diagrams). Fig. 1 illustrates the key mappings.

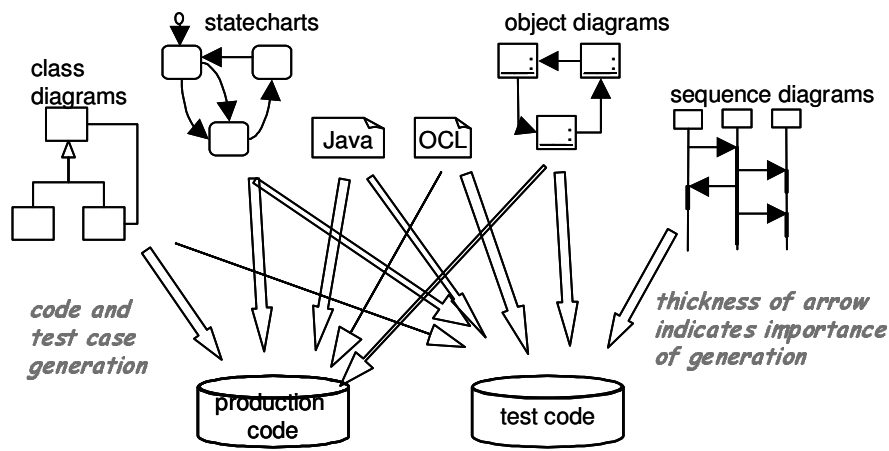

**Fig. 1.** Mapping of UML-models to code and test code.

As a consequence of the various possible forms of model use, we will identify the notion of diagram and model. Thus a *model* is a coherent piece of information, denoted in a diagrammatic or textual notation that describes an abstraction of the desired system. Multiple models can describe various aspects of the system. We also allow models to be composed of sub-models, as is the case with test models. This is in slight contrast to approaches, where a model is a virtual something in a tool and the user manipulates it indirectly through (diagrammatic) views. The latter approach has shown some difficulties when using models for test case definition.

## 3 Testing Strategies

Not only [Bin99,Mye79] show that there is a huge variety of testing strategies and testing goals. While tests in the small usually try to identify faults, tests suites and coverage metrics can be used to estimate the quality of the system in terms of absence of faults. The "success" of a test can therefore be seen twofold, but we follow the general line and define a test to *fail*, when an abnormal behavior (*failure*) shows that there is at least one *fault* in the system. This means the test and the system do not fit together.

Testing can be done manually or automated. The widespread use of JUnit [BG99] shows that the use of automated tests has gained considerable attention in recent years, because it allows to "reuse" tests in form of regression tests on evolving systems without actually knowing what the test does. This allows very small iterations with continuous integration and the use of refactoring techniques [Fow99] to improve the code structure. Automated tests ensure a low defect rate and continuous progress, whereas manual tests would very rapidly lead to exhausted testers. To summarize the characteristics of tests we are aiming at:

- Tests run the system – in contrast to static analyses.
- Tests are automatic to prevent project members to get bored with tests (or alternatively to prevent a system that isn't tested enough)
- Automated tests build the *test data*, run the test and *examine the result* automatically. *Success* resp. *failure of the test* are automatically observed during the test run.
- A *test suite* also defines a system that is running together with the tested production system. The purpose of this extended system is to run the tests in an automated form.
- A test is *exemplaric*. A test uses particular values for the input data, the *test data*.
- A test is *repeatable* and *determined*. For the same setup the same results are produced.

In particular the last point is tricky to achieve, when the *system under test* (SUT) is distributed or has side effects. Specific care has to be taken to deal with these situations. Faults may also occur without being observable, because they are hidden in the SUT or cannot be traced in a distributed system. That is why the TTCN standard [TTCN92] also allows "inconclusive" and "none" as test results.

We instead strongly demand that systems need to be built in a way that testing can be properly achieved. At least in the domain of object-oriented software systems this is a realistic demand. After discussing the structure of a model-based test in the following section, we will discuss, how object-oriented systems should be designed to assist automated testing.

## 4 Model-based Testing

The use of models for the definition of tests and production code can be manifold:
- Code or at least code frames can be generated from a design model.
- Test cases can be derived from an analysis or design model that is not used/usable for constructive generation of production code. For example behavioral models, such as statecharts, can be used to derive test cases that cover states, transitions or even larger subsets of its paths.
- The models itself can be used to describe test cases or at least some part thereof.

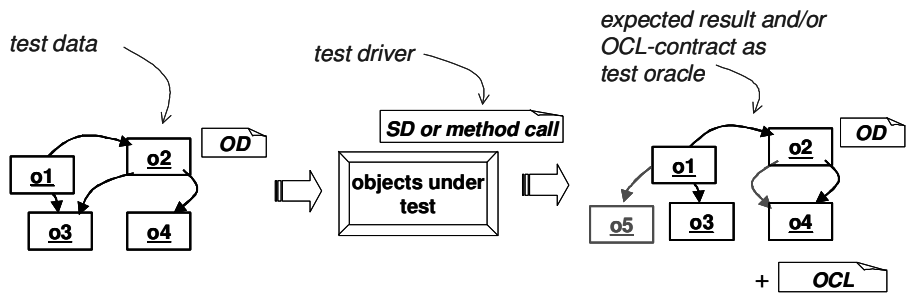

**Fig. 2.** Structure of a test modeled with object diagrams (OD), sequence diagram (SD) and the Object Constraint Language (OCL).

The first two uses are already discussed e.g. in [Rum02] and [BL01]. Therefore, in this section we concentrate on the development of models that define tests. A typical test, as shown in Fig. 2 consists of a description of the test data, the *test driver* and an *oracle* characterizing the expected test result. In object-oriented environments, test data can usually be described by an object diagram (OD). The object diagram in Fig. 3 shows the necessary objects as well as concrete values for their attributes and the linking structure.

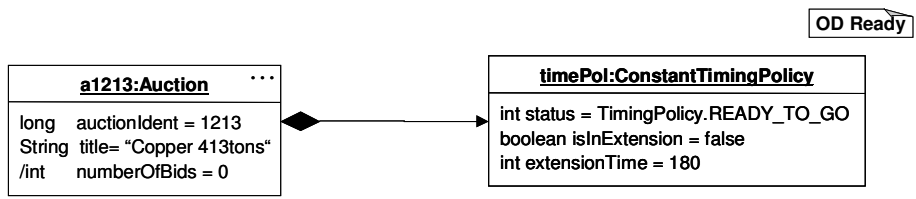

**Fig. 3.** Object diagram (OD) describing the test data as a particular situation in an online auction, that has not yet started (see the full example in [Rum03,Rum03b]).

The test driver can be defined using a simple method call or, if more complex, modeled by a sequence diagram (SD). An SD has the considerable advantage that not only the triggering method calls can be described, but it is possible to model desired interactions and check object states during the test run. For this purpose, the Object Constraint Language (OCL, [WK98]) is used. In the sequence diagram in Fig. 4, an OCL constraint at the bottom ensures that the new closing time of the auction is set to the time when the bid was submitted (bid.time) plus the extension time to allow competitors to react (the auction system using this structure is partly described in [Rum03, Rum03b]).

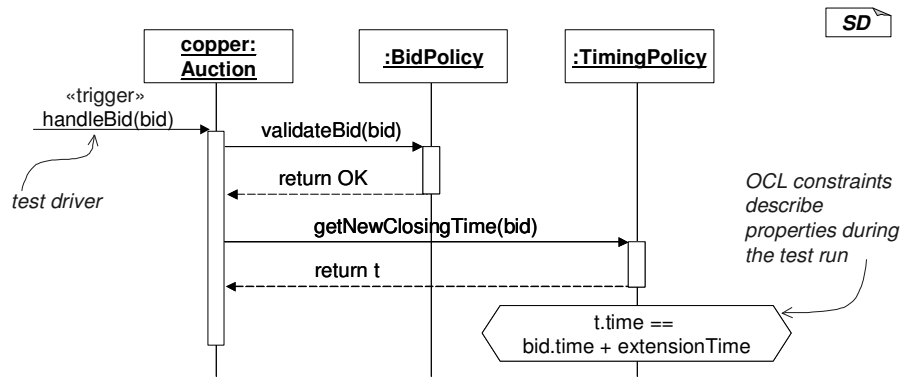

**Fig. 4.** A sequence diagram (SD) describing the trigger of a test driver and predicts some interactions as well as an OCL property that holds at that point in the test.

It is effective to model the test oracle using a combination of an object diagram and reuse globally valid OCL properties. The object diagram in this case serves as a property description and can therefore be rather incomplete, just focusing on desired effects. The OCL constraints can also be taken from the set of general invariants or can be defined as specific properties. In practice, it turns out that there is a high amount of reuse possible through the following techniques:

- Well prepared test data can be reused for many tests. There is often only a handful of *basic test structures* necessary. From those the specifically desired test data can be derived by small adaptations, e.g. replacing a single attribute value or adding a certain object. Having an explicit, graphically depicted basic model of the data structure at hand increases its reusability for specific adaptations.
- Test data can be composed of several object diagrams, describing different parts of the data.
- Test oracles can be defined using a combination of an object diagram and reuse globally valid OCL properties. The resulting object diagram can be rather small, describing deltas only and can be derived from the test data diagram. The OCL properties can be reused as they are usually globally valid invariants.

As already mentioned, being able to use the same, coherent language to model the production system and the tests allows for a good integration between both tasks. It allows the developer to immediately define tests for the constructive model developed. It is therefore feasible that in a kind of "test-first modeling approach" the test data in form of possible object structures is developed before the actual implementation.

## 5 Test Pattern

In the last few years a number of Agile Methods have been defined that share a certain kind of characteristics, described in [AM03]. Among these Extreme Program-ming (XP) [Bec99] is the most widely used and discussed method. One of the most important XP characteristics is that it uses automated tests at all stages. Practical experience shows that when this is properly done, the defect rate is considerably low [RS02]. Furthermore, automation allows to repeat tests continuously in form of regression tests. Thus the quality of the result is ensured through strong emphasis on testing activities, ideally on development of the tests before the production code ("test first approach" [LF02]). When using UML models for test design the development project should become even more efficient.

However, practical experience shows that there are a number of obstacles that need to be overcome to enable model-based testing. In particular, there are object-oriented architectures that exhibit serious problems that prevent tests. It is therefore important to identify those problems and offer appropriate and effective solutions. In the remainder of this section, we provide several solutions for a number of problems that typically occur and that we also experienced e.g. in the online auction system. These solutions are defined in form of applicable test patterns similar to the design patterns of [GHJV94]. Indeed, some of the test patterns are based on design patterns, such as singleton, adapter or factory to achieve a testable design. Unlike [Bin99] we only provide the essential structure and a short explanation of the patterns in this article and refer to [Rum03] to a more detailed description.

A test pattern description typically consists of several parts, describing intention, how to apply, the resulting structure, example implementations and a discussion of the pros and cons. Often the structure itself appears as a simple concept and it's the method part, describing practical knowledge of its applicability that makes a pattern useful.

**Dummies for the test context**

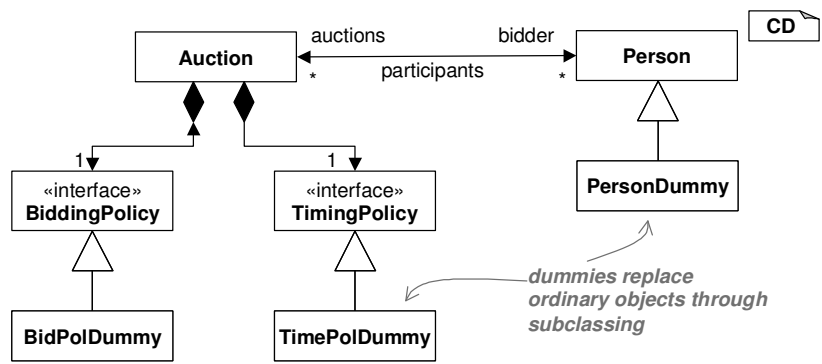

**Fig. 5.** Class diagram showing how dummies are added to the system.

It has become a primary testing technique to use dummies (also "stubs" or "mocks") to replace parts of the system and thus better expose the tested part to the test driver. Only object-oriented concepts, namely inheritance and dynamic binding (also known as polymorphic replacement of objects), comfortably allows to build object structures that are testable with dummies. Fig. 5 shows the principle in one class diagram that allows to isolate an auction object, by replacing its context completely by dummies.

Sometimes a dummy just does nothing, but often it is also necessary to feed back specific values to keep the test going in the desired direction. Practical experience shows that this should normally not be achieved through various dummy-subclasses, but through parameterized dummies, whose behavior during the test can be determined via constructor parameters. This for example allows to predefine and store results of queries given back to the calling SUT just to see what the SUTs reaction will be on that data.

**Remembering interaction and results**

A typical application of a dummy is to prevent side effects that a system otherwise has on the environment. Such side effects may affect files, the data base, the graphical user interface etc. An object responsible for logging activities that provides a method "write" may be replaced by a subclass object (say "LogDummy") where the "write" method simply stores the line to write in a local attribute where it can be examined after the test. Sequence diagrams, however, already allow access to this kind of values during the test. Fig. 6 describes the effect of a test on the log object directly.

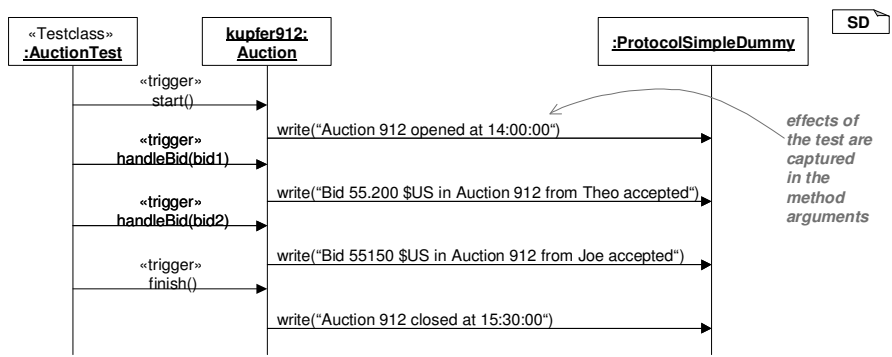

**Fig. 6.** Describing the effects for the log in a sequence diagram.

**Static elements**

As explained two concepts of object-oriented languages, namely inheritance and dynamic binding, allow to setup tests in a form that was not possible in procedural and functional languages. However, typical OO languages also provide concepts that make testing difficult. These are in particular static concepts, such as static attributes, static methods and constructors.

Static attributes are mainly used to allow easy sharing of some global resources, such as the log object or database access. Static attributes, however, should be avoided anyway. If necessary e.g. for access to generally known objects, a static attribute can at least be encapsulated by a static method.

The problem with a static method results from the inability to redefine it for testing purposes. For example if the static method "write" does have side effects, these cannot be prevented during a test. Unfortunately, there are often at least some static methods necessary. We have therefore used the technique shown in Fig. 7 to provide a static interface to the customer and at the same time to allow the effect of the static method to be adaptable, through using an internal delegation to a singleton object that is stored in a static attribute. With proper encapsulation of the initialization of that attribute this is a safe and still efficient technique to make static methods replaceable by dummies without changing their signature. Thus the side effects of static methods can be prevented.

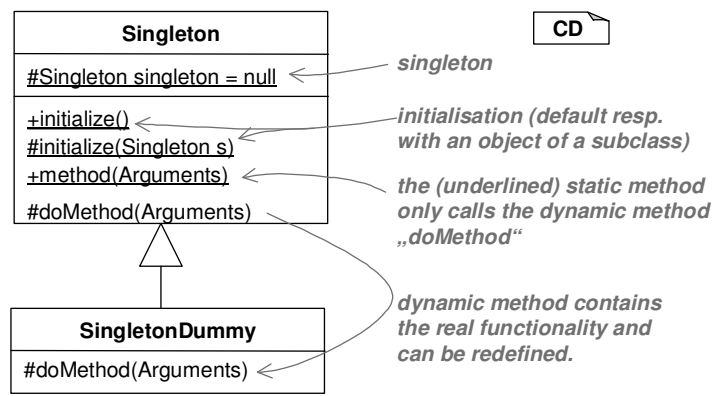

**Fig. 7.** Singleton object behind a static method.

**Constructors**

With respect to testing purposes a constructor shares a number of similarities with static methods, because a constructor is not dynamically bound and can therefore not be redefined. Furthermore, a constructor creates new objects and thus allows the object under test to change its own environment. For example the tested object may have the idea to create its own log object and write the log through it.

The standard solution for this problem is to force any object creation to be done by a factory. So instead "new class(arg)" a method "getClass(arg)" might be called. This method may be static using the approach above to encapsulate the factory object, but still make it replaceable for tests. A factory dummy can then create objects of appropriate subclasses that serve as dummies with certain predefined test behavior.

In practice, we found it useful to model the newly created objects using object diagrams. The factory dummy that replaces the factory then doesn't really create new objects, but returns one of the predefined objects each time it is called. Fig. 8 shows the factory part of a test data structure where three different person objects shall be "created" during the test. The data structure and the attribute values can be described using the same object diagram as is used when describing the basic test data structure. Further advantages are (1) that the newly "created" objects are known by name and can thus easily be checked after the test, even if the objects were disposed during the test and (2) the order of creation can also be checked.

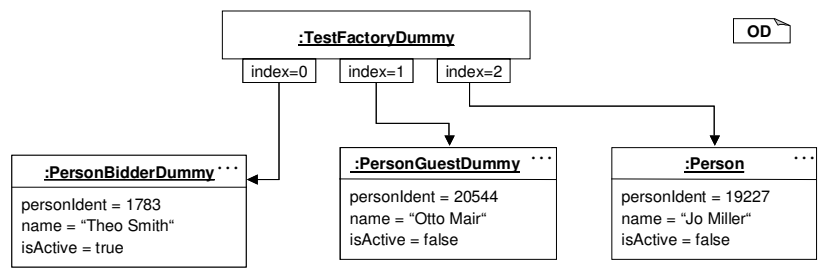

**Fig. 8.** Replacing a constructor by a factory and modeling the factory behavior through a series of objects to be "created" during a test.

**Frameworks and components**

Frameworks are commonly used in object-oriented systems. Most prominent examples are the AWT and Swing from Java, but basic classes, such as the containers partly also belong to that category. Software that uses predefined frameworks is particularly hard to test:

- The control flow of the framework makes tests tricky to run.
- Form and class of newly created objects within the framework is predefined through the used constructors.
- Static variables and methods of the framework cannot be controlled, in particular if they are encapsulated.
- Encapsulation prevents to check the test result.
- Sometimes subclasses cannot be built properly, because the class or methods are declared "final", there is no public constructor, the constructor does have side effects, or the internal control flow is unknown.

Today there doesn't exist a single framework that is directly suited for tests. Such a framework should allow to replace framework objects by self-defined subclass objects and should provide its own default dummy subclasses. Furthermore, the framework should use factories for object creation and give the test developer a possibility to replace these factories. The white-box adaptation principles that frameworks usually provide through subclassing are indeed helpful and sometimes sufficient, but if not, a more general technique, the adapter, is needed to separate application and frame-work. This is a recommended technique for application developers anyway to decouple application and framework and can be reused for improvement of testability as well.

Fig. 9 shows how a JSP "ServletRequest" class is adapted. A "ServletRequest" basically contains the contents of a web form filled by the user in form of pairs (parametername, content). Unfortunately "ServletRequest"-objects can only be created by handling actual requests through the web. Therefore, an adapter is used, which is called "OwnServletRequest".

In an adapter normally simple delegation is used. But, as framework classes are strongly interconnected method calls often require other framework objects as parameters or reveal access to other framework objects. For example the method "getSession()" needs an additional wrapping to return the proper object of class "OwnSession".

This adapter technique allows us to completely decouple application and framework and even to run the application part without the framework as it may be desired in tests. For testing purposes "OwnServletRequestDummy" may now overwrite all methods and use a "Map" to store a predefined set of "user" inputs.

However, there must be noted that this kind of wrapping may need additional infrastructure to ensure that each time "getSession()" is called on the same "ServletRequest", the same corresponding session object is returned. This can be solved through an internal Map from Session to OwnSession that keeps track.

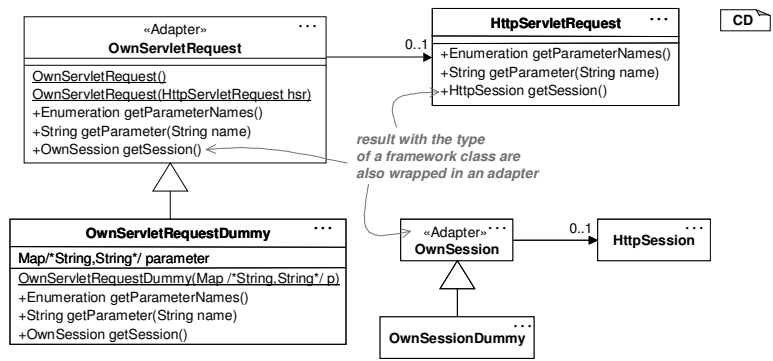

**Fig. 9.** Adapters for framework classes.

So far we have discussed a number of easy to apply, but effective techniques to make object-oriented systems testable. We have used class, sequence and object diagrams to model the test pattern and demonstrated how to use these diagrams to model the test data and dummies. It is an interesting question, how to present such methodological knowledge and its individual parts. Basically the technical principles, such as pattern structure can be formally defined. This has the advantage that at least the structural part of such a pattern can automatically be applied using an appropriate tool. However, the most important part of a test pattern, namely the methical ex-perience cannot be formalized, but needs to be presented to the user in an understandable way. The user needs then to adapt a pattern to his specific situation. Therefore, we have chosen to largely use examples instead of precise descriptions for patterns.

## 6  Test Pattern for Distributed Systems

Testing a distributed system may become hard, as distribution naturally involves concurrent processes with interactions and timeouts thus leading to nondeterministic behavior. Furthermore, it is normally not possible to stop the system and obtain a global system state for consistency checking. There exists a variety of approaches to deal with these problems, in particular in the hardware and embedded systems area. Through the distribution of web services in particular in the E-Commerce domain, it becomes increasingly important to be able to deal with distributed object systems in this domain as well. In our example application, the online auction system, timing and distribution are very important, as auctions last only a very restricted time (e.g. one hour) and in the final phase bids are submitted within seconds. Therefore, it is necessary that auctions are handled synchronously over the web.

The test patterns discussed in this section tackle four occurring problems: (1) simulation of time and progress, (2) handling concurrency through threads, (3) dealing with distribution and (4) communication. As already mentioned, the test patterns concentrate on functional tests. Additional effort is necessary to test quality of service attributes, such as throughput, mean uptime, etc. The proposed techniques have already been used in other approaches, the novelty basically comes from the combination of modeling techniques and these concepts in form of methodical test patterns.

**Handling time and progress**

An online auction usually takes about one hour. However, a single test may not take that time, but needs to be run in milliseconds, as hundreds of tests shall finish quickly. So it is necessary to simulate time. This becomes even more important, when distributed processes come into play that do not agree on a global time as is usually the case in the internet.

Thus instead of calling the time routine of the operating system directly, an adapter is used. The adapter can be replaced by a parameterized dummy that allows us to freely set time. For many tests, a fixed time is sufficient, for tests of larger series of behaviors, however, it is also necessary that progress happens. Thus in the time pattern two more concepts can be established: First, each query of the current time increases time for one tick. Second, we use explicit time stamps on sequence diagrams to adapt time during the test. The time stamps as shown in Fig. 10 therefore correspond to statements that update the timing dummy. This active use of time stamps contrasts other approaches, where a passive interpretation regards a time stamp as maximum durations that a signal may take. The principle used here to simulate time also allows to simulate the behavior of times that trigger certain events regularly or after timeouts.

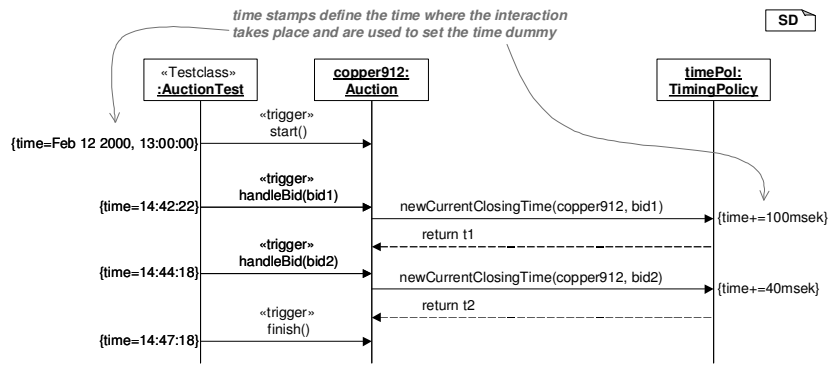

**Fig. 10.** A sequence diagram with time stamps to describe the progress of time.

**Concurrency using threads**

Concurrency within one processing unit is often used to increase reactivity and to delegate regularly occurring tasks to specific units. In web applications, threads deal with polling of TCP/IP-data from sockets and with GUI interactions. However, those threads are normally encapsulated in the frameworks and use "callbacks" to the application code to handle a request. For a functional test of this type of concurrency it is necessary to simulate these callbacks. This can be done by defining a fixed scheduling for callbacks to obtain the necessary determinism and repeatability. Fig. 11 shows a test driver in form of a sequence diagram, where the driving object submits several requests to a number of objects that are normally running within different threads.

This approach only works for sequential calls and therefore doesn't test whether the simulated threads would behave similar, if running in parallel resp. a machine scheduled interleaving. Thus we just do functional tests. On one hand interleaving can be checked through additional stress tests and reviews. On the other hand Java e.g. provides a synchronization concept that if properly used is a powerful technique to make programs thread-safe. In practice concurrency problems have been considerably reduced since the concept of thread-safeness was introduced. In more involved situations, where interaction between two active threads is actually desired and therefore shall be tested, it might be necessary to slice methods into smaller portions and do a more fine grained scheduling. However, the possibilities of interactions easily explode and efficient testing strategies are necessary. It is also possible to set up test drivers that run large sets of generated tests to explore at least a part of the interaction space.

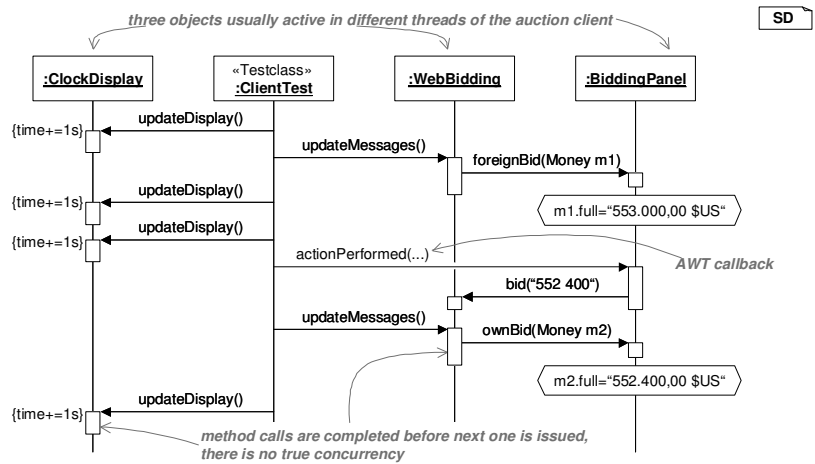

**Fig. 11.** A test driver schedules calls to objects that normally reside in different threads.

In an application where the developers define threads on their own, these threads usually have the form of a loop with a regularly repeated activity and a sleep statement. If not, they usually can and for applicability of the test pattern also should be reformulated in such a form. The repeating activity then can easily be added to the test scheduling, whereas the Thread object itself should be so simple, that a review is sufficient to ensure its correctness.

In Java the Thread class provides additional functionality, such as a join or termination of threads which causes additional effort to simulate. As some of those methods cannot be redefined in subclasses it might be necessary to use an adapter.

**Distributed Systems**

Based on the test pattern defined so far, it now becomes feasible to test distributed systems. Real or at least conceptually distributed systems have subsystems with separated storage and enforce explicit communication. With CORBA, DCOM, RMI or even plain socket handling there is a variety of communication techniques available.

Of course it is possible to run distributed tests, but it is a lot more efficient to simulate the distribution within one process. Again this technique only works for tests of the functionality. One cannot expect to get good data on reactivity and efficiency of the system when several subsystems are mapped into one process. As each object in the distributed system resides in exactly one part, we introduce a new tag, called *location* that allows to model in the test, where the object resides. Fig. 12 shows a test driver with an interleaving of activities in distributed locations.

To simulate a distributed system it is necessary to ensure that the distributed threads are mapped into one process in such a way that no additional interactions occur. But interactions usually occur when static state is involved, because e.g. static attributes can be globally accessed. In a distributed system every subsystem had its own static attribute, after the mapping only one attribute exists. Our encapsulation of static attributes in singleton objects, however, can easily be adapted to simulate a multiple static attribute. Actually the delegation mechanism explained earlier is extended to use a map from location to attribute content instead of a single attribute. The location is set by the test driver accordingly thus allowing to distinguish the respective context of each tested object. This for example allows to handle multiple logs, etc.

The location tag is therefore crucial to setup virtually distributed systems and run them in an interleaved manner in such a way that they believe they run on their own.

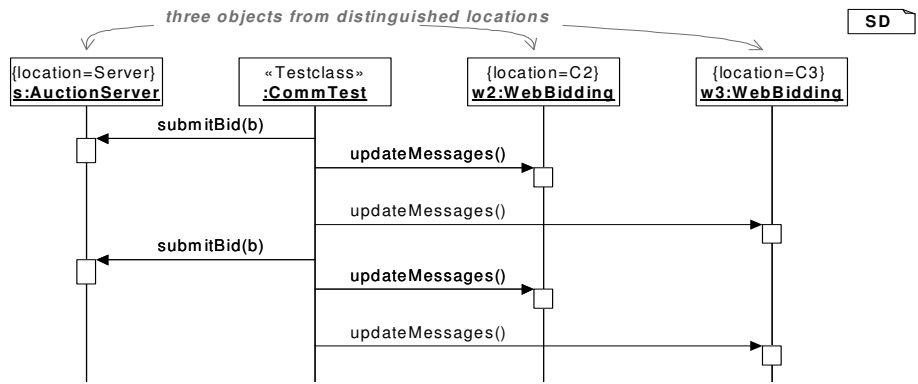

**Fig. 12.** A test driver schedules calls to objects in different locations.

The virtually distributed system, however, gives us an interesting opportunity: It allows to stop the system during the run and check invariants and conditions across subsystem borders by simply adding globally valid OCL-constraints to the sequence diagram that drives the test run. To be furthermore able to talk about local invariants, we have extended the OCL in [Rum03] to also allow localized constraints.

**Distributed communication**

The remaining problem for tests of distributed systems is to simulate communication in an efficient way. The standard technique here is to build layers of respective communicating objects and use proxies (stubs) on each layer where appropriate. If for example CORBA is used, we build an adapter system around the CORBA API to encapsulate it in the same way as for ordinary frameworks. Thus replacement of the communication part through a dummy becomes feasible.

In Fig. 13 we see two object diagrams showing layers of a subset of the communication mechanism that directly deals with sockets in Java. A bid arrives at the AuctionServerProxy in the client. It is transformed into a string and transferred via the MessageHandleProxy, the BufferedWriter and the URLConnection to the socket on the server side. There a thread that resides in the HttpConnection sleeps until a string is received on the socket. The received strings is transferred to the actual MessageHandler that un-marshalls the object into the original bid and gives it to the actual auction server.

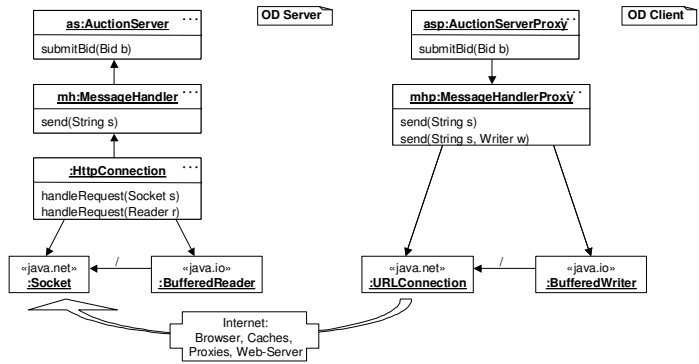

**Fig. 13.** The layers of communication objects in the original system.

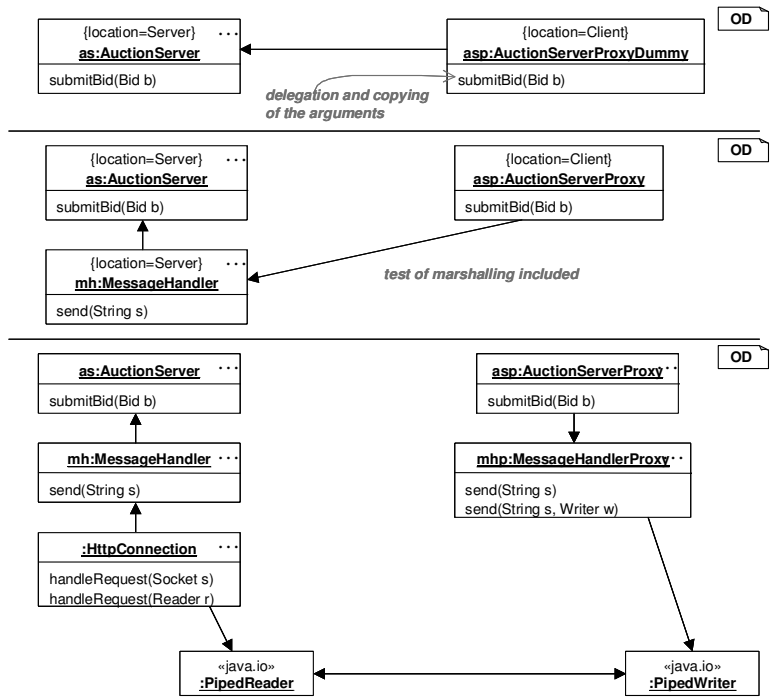

**Fig. 14.** Three shortcuts for the communication layer.

The trick is that both proxies on the right hand side resemble the same signature by sharing the same interface as their real counterparts on the left hand side. Therefore in a test we may simply shortcut the communication structure. Depending on the objects, we want to test, we might shortcut at the AuctionServer-layer already or go as deep as the Reader/Writer-pair. Fig. 14 shows three variants of possible connections. In the first two configurations it is important to model the location of each object, because the test generator needs to ensure the location is changed, when a client object calls a server object. In the third configuration it is unnecessary, as a transfer of a bid now consists of two parts. First, the bid is submitted at the AuctionServerProxy on the client side and stored at the PipedReader and then the HttpConnection is activated in a second call on the server side.

In the preceeding two sections, we have discussed a number of test patterns using models to describe the basic structure. The test patterns on one hand allow us to actually define functional tests for almost any kind of object-oriented and in particular distributed system in a systematic way. On the other hand the used examples show how easy it can be to define and understand test setups that are based on models. These models are a lot more compact and can more easily be developed, read and understood than code. Increased usability of these models for several development stages becomes feasible, because of a better understanding what these models can be used for. Therefore, model-based development as proposed by the MDA-approach [OMG01] becomes applicable.

## 7 Model Evolution using Automated Tests

Using models for test and application development is only one side of the medal. Automated testing is the primary enabler for an evolutionary approach of developing systems. Therefore, in this section, we give a sketch of how model-based, automated testing, and model evolution fit together.

In the development approach sketched so far, an explicit architectural design phase is abandoned and the architecture emerges during design. Architectural shortcomings are resolved through the application of refactoring techniques [OJ93,Fow99]. These are transformational techniques to evolve a system in small, systematic steps to enhance its structure. The concept isn't new (see [PR03] for a discussion), but through availability of tools and its embedding in XP [Bec99], transformational development now becomes widely used.

Nowadays, it is expected that the development and maintenance process is capable of being flexible enough to dynamically react on changing requirements. In particular, enhanced business logic or additional functionality should be added rapidly to existing systems, without necessarily undergo a major re-development or re-engineering phase. This can be achieved at best if techniques are available that systematically evolve the system using transformations. To make such an approach manageable, the refactoring techniques for Java [Fow99] have proven that a comprehensible set of small and systematically applicable transformation rules seems opti-mal. Transformations, however, cannot only be applied to code, but to any kind of model. A number of possible applications are discussed in [PR03].

Having a comprehensible set of model transformations at hand, model evolution becomes a crucial step in software development and maintenance. Architectural and design flaws can then be more easily corrected, superfluous functionality and structure removed, structure for additional functionality or behavioral optimizations be adapted, because models are more abstract, exhibit higher-level architectural and design information in a better way.

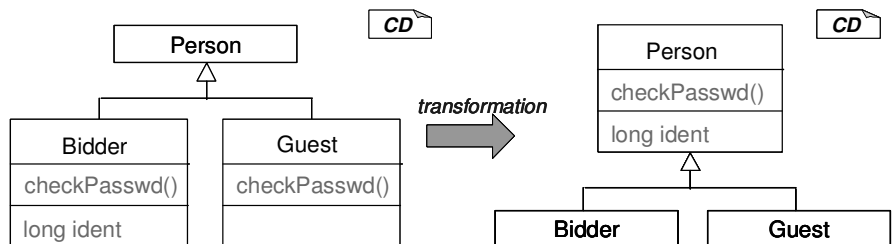

**Fig. 15.** Two transformational steps moving an attribute and a method along the hierarchy.

Two simple transformation rules for a class diagram are shown in Fig. 15. The figure shows two steps that move a method and an attribute upward in the inheritance hierarchy at once. The upward move of the attribute is accompanied by the only context condition that the other class "Guest" does not have an attribute with the same name yet. In contrast, moving the method may be more involved. In particular, if both existing method bodies are different, there are several possibilities: (1) Move up one method implementation and have it overridden in the other class. (2) Just add the methods signature in the superclass. (3) Adapt the method implementations in such a way that distinguishing parts are factored out into other sub-methods and the remainder of the method bodies is identical in both methods.

Many of the necessary transformation steps are as simple as the upward move of an attribute. However, others are more involved and their application comes with a larger set of context conditions. These of course need automated assistance. The power of these rather simple and manageable transformation steps comes from the possibility to combine them and evolve complex designs in a systematic and traceable way.

Following the definition on refactoring from [Fow99], we use transformational steps for structure enhancement that do not affect "externally visible behavior". For example both transformations shown in Fig. 15 do not affect the external behavior if made properly.

By "externally visible behavior" Fowler in [Fow99] basically refers to behavioral changes visible to the user. This can be generalized by introducing an abstract "system border" that may also act as interface to other systems. Furthermore, in a hierarchically structured system, we may enforce behavioral equivalence for "subsystem borders" already. It is therefore necessary to explicitly describe, which kind of behavior is regarded as externally visible. For this purpose tests are the appropriate tech-nique to describe behavior, because (1) tests are already available as result of the development process and (2) tests are automated which allows us to check the effect of a transformation through inexpensive, automated regression testing.

A test case thus acts as an "observer" of the behavior of a system under a certain condition. This condition is also described by the test case, namely through the setup, the test driver and the observations made by the test. Fig. 16 illustrates this situation.

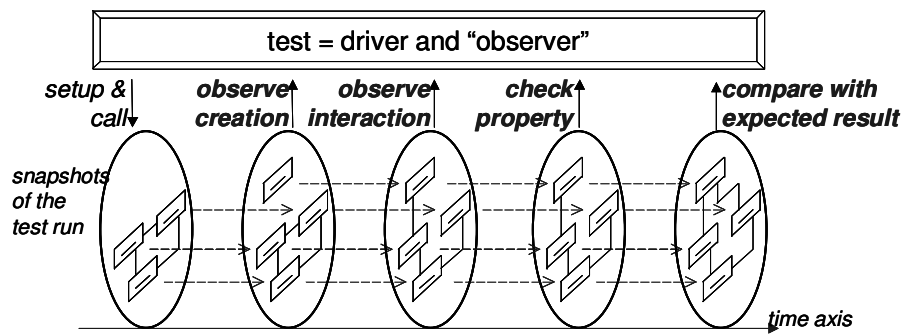

**Fig. 16.** A test case acts as observation.

Fig. 16 also shows that tests do not necessarily constrain their observation to "externally visible behavior", but can make observations on local structure, internal interactions or state properties even during the system run. Therefore, we distinguish "internal" test that evolve together with the transformed system and "external" tests which need to remain unchanged, because they describe external properties of the system.

*Unit and integration tests* focus on small parts of the system (classes or subsystems) and usually take a deep look into system internals. These tests are usually transformed together with the code models. Unit and integration tests are usually provided by the developer or test teams that have access to the systems internal details. Therefore, these are usually "glass box tests".

*Acceptance tests*, instead, are "black box" tests that are provided by the user (although again realized by developers) and describe external properties of the system. These tests must be a lot more robust against changes of internal structure. Fig. 17 illustrates a diagram that illustrates how an observation remains invariant under a test. To achieve robustness, acceptance tests should be modeled against the published interfaces of a system. In this context "published" means that parts of the system that are explicitly marked as externally visible and therefore usually rather stable. Only explicit changes of requirements lead to changes of these tests and indeed the adaptation of requirements can very well be demonstrated through adaptation of these test models followed by the transformations necessary to meet these tests afterwards in a "test-first-approach".

To increase stability of acceptance tests in transformational development, it has proven useful to follow a number of standards for test model development. These are similar to coding standards and have been found useful already before the combination with the transformational approach:
- In general an acceptance test should be abstract, by not trying to determine every detail of the tested part of the system.
- A test oracle should not try to determine every part of the output and the resulting data structure, but concentrate on important details, e.g. by ignoring uninteresting objects and attribute values (e.g. in object diagrams and OCL constraints).
- OCL property descriptions can often be used to model a range of possible results instead of determining one concrete result.
- Query-methods can be used instead of direct attribute access. This is more stable when the data structure is changed.
- It should not be tried to observe internal interactions during the system run. This means that sequence diagrams that are used as drivers for acceptance tests concentrate on triggers and on interactions with the system border.
- Explicitly published interfaces that are regarded as highly stable should be introduced and acceptance tests should focus on these interfaces.

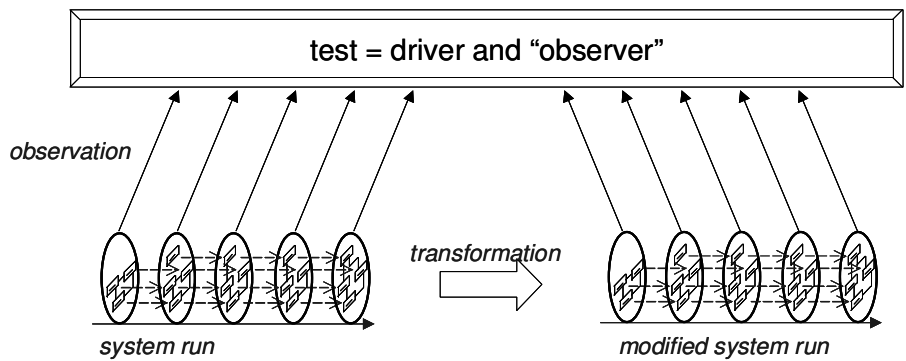

**Fig. 17.** The transformed system model is invariant under a test observation.

## 8   Conclusions

The proposal made in this paper is part of a pragmatic approach to model-based software development. This approach uses models as primary artifact for requirements and design documentation, code generation and test case development and includes a transformational technique to model evolution for efficient adaptation of the system to changing requirements and technology, to optimize architectural design and fix bugs. To ensure the quality of such an evolving system, intensive sets of test cases are an important prerequisite. They are modeled in the same language, namely UML, and thus exhibit a good integration and allow us to model system and tests in parallel.

The paper demonstrates that it is feasible to use various kinds of models to explicitly define automated tests. For use in object-oriented systems, however, the design of the system has to some extent to be adapted in order to allow testable systems. A series of basic and enhanced test patterns lead to a better testable design. In particular test patterns for distributed systems are a necessary prerequisite to allow testability.

However, there are some obstacles for the proposed approach. (1) Currently, tool assistance is still in its infancy. (2) More experience is needed to come up with effective testing techniques in the context of model evolution, which must also involve coverage metrics. (3) These new techniques, namely an executable sub-language of the UML as well as a lightweight methodological use of models in a development process are both a challenge to current practice in software engineering. They exhibit new possibilities and problems. Using executable UML allows to program in a more abstract and efficient way. This may finally downsize projects and decrease costs. The free resources can alternatively be used within the project for additional validation activities, such as reviews, additional tests or even a verification of critical parts of the system.

Therefore, we can conclude that techniques such as model-based development, model evolution and test-first design will change software engineering and add new elements to its portfolio.

**Acknowledgements**

I would like to thank Markus Pister, Bernhard Schätz, Tilman Seifert and Guido Wimmel for commenting an earlier version of the paper as well as for valuable discussions. This work was partially supported by the Bayerisches Staatsministerium für Wissenschaft, Forschung und Kunst and through the Bavarian Habilitation Fellowship, the German Bundesministerium für Bildung und Forschung through the Virtual Software Engineering Competence Center (ViSEK).